\documentclass[12pt]{article}

\usepackage{scicite}
\usepackage{color}
\usepackage{graphicx}

\usepackage{times}
\usepackage{ulem}
\usepackage{amssymb,amsmath}

\newenvironment{sciabstract}{%
\begin{quote} \bf}
{\end{quote}}

\newcounter{lastnote}

\title{KOI-126: A Triply-Eclipsing
Hierarchical Triple \\ with Two Low-Mass Stars}

\author
{Joshua A. Carter,$^{1, 2\ast}$ Daniel C. Fabrycky,$^{2,3}$ Darin Ragozzine,$^{1}$ \\
Matthew J. Holman,$^{1}$ Samuel N. Quinn,$^{1}$ David W. Latham,$^{1}$  \\
Lars A. Buchhave,$^{1,4}$ Jeffrey Van Cleve,$^{5,8}$ William D. Cochran,$^{6}$ \\
Miles T. Cote,$^{5}$ Michael Endl,$^{6}$  Eric B. Ford,$^{7}$  \\
Michael R. Haas,$^{5}$ Jon M. Jenkins,$^{5, 8}$ David G. Koch,$^{5}$  Jie Li,$^{5, 8}$ \\ 
Jack J. Lissauer,$^{5,9}$   Phillip J. MacQueen,$^{6}$ Christopher K. Middour,$^{5,13}$ \\
Jerome A. Orosz,$^{10}$ Jason F. Rowe,$^{5,8,11}$ Jason H. Steffen,$^{12}$ William F. Welsh$^{10}$\\
\\
\normalsize{$^{1}$Harvard-Smithsonian Center for Astrophyics, 60 Garden Street, Cambridge, MA 02138, USA.}\\
\normalsize{$^{2}$Hubble Fellow.}\\
\normalsize{$^{3}$UCO/Lick, University of California, Santa Cruz, CA 95064, USA.}\\
\normalsize{$^{4}$Niels Bohr Institute, Copenhagen University, DK-2100 Copenhagen, Denmark.}\\
\normalsize{$^{5}$NASA Ames Research Center, Moffett Field, CA 94035, USA.}\\
\normalsize{$^{6}$University of Texas, Austin, TX 78712, USA.}\\
\normalsize{$^{7}$University of Florida, Gainesville, FL 32611, USA.}\\
\normalsize{$^{8}$ SETI Institute, Mountain View, CA 94043, USA.}\\
\normalsize{$^{9}$Stanford University, Stanford, CA 94305, USA.}\\
\normalsize{$^{10}$San Diego State University, San Diego, CA 92182, USA.}\\
\normalsize{$^{11}$NASA Postdoctoral Program Fellow. }\\
\normalsize{$^{12}$Fermilab Center for Particle Astrophysics, Batavia, IL 60510, USA.}\\
\normalsize{$^{13}$Orbital Sciences Corp.}\\
\\
\normalsize{$^\ast$To whom correspondence should be addressed; E-mail:  jacarter@cfa.harvard.edu.}
}

\date{}

\begin{document}

\baselineskip24pt

\maketitle 

\newcommand{\ron}{\textcolor{red}}

\begin{sciabstract}
  The Kepler spacecraft has been monitoring the light from 150,000 stars in its primary quest to detect transiting exoplanets.  Here we report on the detection of an eclipsing stellar hierarchical triple, identified in the Kepler photometry. 
 KOI-126~(A,~(B, C)), is composed of a low-mass binary (masses $M_B = 0.2413\pm0.0030$ $M_\odot$, $M_C = 0.2127\pm0.0026$ $M_\odot$; radii $R_B = 0.2543\pm0.0014$ $R_\odot$, $R_C = 0.2318\pm0.0013$ $R_\odot$; orbital period $P_1 =  1.76713\pm0.00019$ days) on an eccentric orbit about a third star (mass $M_A = 1.347\pm0.032$ $M_\odot$; radius $R_A = 2.0254\pm0.0098$ $R_\odot$; period of orbit around the low-mass binary $P_2 = 33.9214\pm0.0013$ days; eccentricity of that orbit $e_2 = 0.3043\pm0.0024$).   The low-mass pair probe the poorly sampled fully-convective stellar domain offering a crucial benchmark for theoretical stellar models.  
  \end{sciabstract}

The Kepler mission's primary science goal is to determine the frequency of Earth-size planets around Sun-like stars.  To accomplish this, thousands of stars are monitored to detect the characteristic dimming of star light associated with a planet passing in front of its host ({\it 1}). 

The Kepler observing scheme lends itself not just to planetary transits but, additionally, to the detection of the analogous eclipses in stellar multiple systems ({\it 2}).  We report on one such hierarchical system consisting of two closely-orbiting low-mass stars (KOI-126~B, C; $P_1 \approx 1.767$ days) that are together orbiting with a longer period ($P_2 \approx 33.92$ days) a more massive and much more luminous star (KOI-126~A).   The two low-mass stars double the number of fully-convective stars with mass and radius determinations better than a few percent; the stars in the eclipsing binary CM Draconis ({\it 3--5}) are the remaining entries in this inventory.  

The KOI-126 system is oriented such that all three components and the Kepler spacecraft lie near a common plane and, consequently, eclipses among all three stars are observed.   The unique variety of these eclipses reveal complex, information-laden dynamical content.   As a consequence, the absolute system parameters for all three stars may be accurately determined from the photometry alone and are immune to the biases  that plague the traditional study of low-mass eclipsing systems (e.g., as a result of stellar spots on the low-mass stars, {\it 6}).  

 The Kepler photometry, reported here, spans 418 days of near continuous observation of KOI-126 (KIC 5897826, 2MASS J19495420+4106514, ${\rm Mag}_{R} = 13.044$) .  The first 171 days consist of successive 29.426 min exposures, whereas the remaining 247 days were observed at both the 29.426 min (or long) cadence and at the 58.85 sec (or short) cadence (Fig. S1 shows the long cadence time-series).  We used the short cadence data ({\it 7}) exclusively in our determination of system parameters.

KOI-126 was originally identified by the Kepler data pipeline ({\it 8, 9}) as a planetary candidate.  The suspected transit events (Fig. 1) were found to be unusual upon closer inspection.  The 1--3\% decrements in the relative flux, occurring approximately every 34 days, exhibit a variable eclipse profile.  At some epochs, two resolved transits are seen; at others, the transits are nearly simultaneous.  The transit-like events have durations that vary substantially across the observed epochs and show asymmetric features about mid-eclipse, indicative of accelerations transverse to the line-of-sight. 

The periodic, superposed transits in the Kepler light curve are most readily explained as the passage of a close (or inner) binary (KOI-126~B, C) across the face of a mutually orbited star (KOI-126~A).  At the start of each passage, KOI-126~B and C  are at a unique phase in their binary orbit, yielding a unique transit route and light curve shape.  To first order, the duration of transit is shorter or longer depending on whether the motion of a given component of the inner binary is prograde (shorter duration) or retrograde (longer duration) relative to the orbit of the inner binary center-of-mass.  The short-timescale orbital motion of the inner binary accounts for the apparent accelerations.

Preliminary modeling of the Kepler light curve predicted the secondary passage of the inner binary behind KOI-126~A (shown in the bottom two time-series in Fig. 1).   A Box-Least-Square algorithm ({\it 10}) search of the Kepler light curve, excluding these secondary and transit events, revealed the relatively shallow eclipses between KOI-126~B and C, occurring every $\approx 0.88$ days (Figs. S2, S3).   Based upon these detections, KOI-126~B and C were inferred to be each less luminous by a factor of $\sim$3,000--5,000, as observed in the wide Kepler bandpass ({\it 1}), than KOI-126~A. Both the eclipses of the inner binary pair and occultations of that pair by KOI-126~A were not observed to be strictly periodic as they were absent in the data for long stretches of time.  The long cadence event near $t = 2,455,069.113$ (BJD) features the alignment of all three objects along the line-of-sight, resulting in a short brightening in the light curve ({\it 11}).  

A periodogram of the light curve, after removing eclipse events and correcting for instrumental systematics, shows an $\approx$17 day modulation with a relative amplitude of $\approx$500 parts-per-million (Fig. S4 plots a representative sample of this variation).  
It is likely associated with the rotation of KOI-126~A (see Supporting Online Material, SOM).

In addition to the Kepler photometry, we collected sixteen spectra of KOI-126  over 500 days (SOM).   The spectra only showed features associated with a single star, KOI-126~A.  The primary goal of these observations was to acquire precision radial velocity (RV) measurements of KOI-126~A (Fig. S5, Table S1).  An ancillary goal of the spectroscopic study was to fit model spectra to the composite spectra for KOI-126~A in order to determine stellar parameters (SOM, Table 1).  From this analysis, we found that KOI-126~A is metal rich, relative to the Sun ([Fe/H] = $0.15\pm0.08$), and has an effective temperature of $5875\pm100$ K and a stellar surface gravity $\log g_A = 3.94\pm0.14$.  

We used a full dynamical-photometric model to explain the data (SOM).  Newton's equations of motion along with a general relativistic correction to the orbit of the inner binary ({\it 12--14}) were integrated to determine the positions and velocities of the bodies at a selected time.   Each individual star's position was then corrected to account for the finite speed of light ({\it 11}).   These corrected positions, along with the absolute object radii and relative flux contributions, were used to calculate the combined flux.

The model was fitted to the short cadence Kepler data by performing a least-squares minimization (SOM).  Only the data shown in Fig. 1 were utilized in the fit.  While not used in the fit, the long cadence data are nearly exactly matched when using the best-fit parameters (Fig. S6).  Subject only to the short cadence photometric data, we determine the individual masses and radii with fractional uncertainties less than $10\%$ and $3\%$, respectively.  

We included the RV data for KOI-126~A in a subsequent fit (Table 1).  The best-fit parameters are identical to those found using photometry alone;  although, masses were determined to better than $3\%$ and radii to better than $0.5\%$. The inner binary orbit is nearly circular and inclined by $\approx9^\circ$ relative to the outer binary orbit.

The photometric data could not be fit by assuming fixed Keplerian orbits for the inner and outer binaries.  This is due to the relatively rapid variation of the Keplerian orbital elements (such as the orbital inclination or the eccentricity) as a result of the gravitational interaction between the three stars (Fig. S7) ({\it 12, 15}).  The observational evidence of this periodicity is most easily seen with the rapid circulation of the ascending node of the inclined inner binary: the inner binary orbit precesses, like a spinning top, every $\sim 1000$ days in response to the gravity of KOI-126~A. This precession explains the occasional absence of the eclipses between KOI-126~B and C.  Similarly, the oscillation of the outer binary inclination, compounded with the measured eccentricity, explains the sparsity of occultations of the inner binary by KOI-126~A.  

Our final dynamical model did not include the effects of stellar spin, tidal distortion, or frictional dissipation because the data did not demand them.  Nevertheless, we did investigate these contributions by extending our numerical model to include parameterized forces appropriate for each effect (SOM).   The only relevant force on the observed timescale, assuming plausible rotation and dissipation rates, is caused by the mutual tidal distortion of KOI-126~B and C.   The controlling parameter for this force is the internal structure (or the apsidal) constant, $k_2$, which is intimately related to the interior stellar density profile and provides an important constraint on stellar models  ({\it 16, 17}).  There are no reliable constraints on $k_2$ for stars with masses similar to KOI-126~B or C. For a fully-convective star, approximated as a polytrope with index $n=1.5$, $k_2$ is estimated to be as large as $0.15$ ({\it 18}).  

Our data constrain $k_2 < 0.6$ at 95\% confidence, assuming that KOI-126~B and C have equivalent apsidal constants.  Although the current constraints on $k_2$ are modest, a fit to the predicted Kepler light curve over the remainder of the nominal mission (3.5 years in total, with 25 more transit-like events) demonstrates that $k_2 \sim 0.1$ will be measured with a relative precision of $\approx 1\%$.   In addition, masses and radii will be determined to better than $0.1\%$. 

The masses of KOI-126~B and C lie below the threshold for having fully convective interiors (less than $\approx0.3 M_\odot$). It has been suggested ({\it 5, 19--22}) that the large (10--15\%) disagreements seen between model predictions ({\it 23, 24}) and measured radii for low-mass stars were confined to stars in close binaries outside of this convective domain.  A sample of low-mass stars with dynamically-measured properties validates this claim ({\it 3, 19--22, 25, 26}); however, there is little reliable information available for stars with masses under $0.3 M_\odot$  (see Fig. 2).  

Previously, CM Draconis ({\it 3--5}) provided the only precise constraints on stellar models for stars below the convective mass boundary.  In its case, theoretical models seemed to underestimate the stellar radii at the 5--7\% level, a disagreement less than that seen with more massive stars but still consistent with increased activity attributed to fast rotation as a result of tidal spin-up ({\it 5}).  In comparison, the radii of KOI-126~B and C are also underestimated by the models; however, this disagreement is smaller (2--5\%; Fig. 2).  

In addition to accurately measured masses and radii, the metallicities and ages of KOI-126~B and C are approximated by the values estimated for KOI-126~A, if we assume all components were co-evolved and formed from the same proto-stellar nebula.  In this case, KOI-126~B and C have a super-solar metallicity ([Fe/H] $= 0.15\pm0.08$) which can be compared to the poorly determined, sub-solar metallicity estimated for CM Dra ({\it 5}).  The enhanced metallicity of KOI-126 relative to solar may partially resolve the discrepancy between the observed and predicted radii of KOI-126~B and C, although we are not aware of any models that properly account for this enhancement.  We compared the mass and radius of KOI-126~A with stellar models in a well-calibrated mass range ({\it 27}) and estimated a system age of $4\pm1$ Gyr. 

We are unable to measure the spin periods of KOI-126~B and C, but, it is likely that synchronization has occurred and that the spin periods are nearly equivalent to the orbital period of the inner binary (SOM).  These spin periods are slower than the expected orbit-synchronized spin periods for CM Dra A and B  ($\approx 1.3$ days) by nearly 0.5 days.  This fact may partially account for the differences in radii between the similar-mass stars CM Dra B and KOI-126 C -- CM Dra B may have increased magnetic activity relative to KOI-126 C owing to its faster rotation ({\it 5}).

\clearpage

{\bf References and Notes}

\begin{enumerate}
\item D. G. Koch {\it et al.}, {\it Astrophys. J.}, {\bf 713}, L79 (2010).										%
\item A. Prsa {\it et al.}, preprint available at http://arxiv.org/abs/1006.2815.                                                                         %
\item C.H. Lacy, {\it Astrophys. J.}, {\bf 218}, 444 (1977).                                                                                                           %
\item T. S. Metcalfe {\it et al.}, {\it Astrophys. J.}, {\bf 456}, 356 (1996).                                                                                       %
\item J. C. Morales {\it et al.}, {\it Astrophys. J.}, {\bf 691}, 1400 (2009).                                                                                   %
\item J.C. Morales, J. Gallardo, I. Ribas, C. Jordi, I. Baraffe, \& G. Chabrier, {\it Astrophys. J}, {\bf 718}, 502 (2010)    %
\item R. L. Gilliland {\it et al.}, {\it Astrophys. J.}, {\bf 713}, L160 (2010).                                                                                        %
\item J. Jenkins {\it et al.}, {\it Astrophys. J.}, {\bf 713}, L87 (2010).                                                                                          %
\item J. Jenkins {\it et al.}, {\it Proc. SPIE}, {\bf 7740} (2010).                                                                                                    %
\item G. Kov{\'a}cs, S. Zucker, T. Mazeh, {\it Astron. Astrophys.}, {\bf 391}, 369 (2002).                                                      %
\item D. Ragozzine, M.J. Holman, {\it Astrophys. J.}, preprint available at http://arxiv.org/abs/1006.3727.                     %
\item S. Soderhjelm, {\it Astron. Astrophys.}, {\bf 141}, 232 (1984).                                                                    			%
\item R.~A. Mardling, D.~N.~C. Lin, {\it Astrophys. J.}, {\bf 573}, 829 (2002).								%
\item M.~H. Soffel, Relativity 																	%
in Astrometry, Celestial Mechanics and Geodesy, XIV, 
Springer-Verlag Berlin Heidelberg New York. (1989)
\item E. Ford, B. Kozinsky, F. Rasio, {\it Astrophys. J.}, {\bf 535}, 385 (2000).                                                                        %
\item Z. Kopal, Dynamics of Close Binary Systems, Reidal, Dordrecht, Holland (1978)                                                     %
\item A. Claret, A. Gim\'{e}nez, {\it Astron. Astrophys.}, {\bf 277}, 487 (1993)                                                                          %
\item R. A. Brooker, T.W. Olle, {\it Mon. Not. R. Astron. Soc.}, {\bf 115}, 101 (1955).                                                              %
\item G. Torres {\it et al.}, {\it Astrophys. J.}, {\bf 640}, 1018 (2006).                                                                                         %
\item M. L\'{o}pez-Morales, {\it Astrophys. J.}, {\bf 660}, 732 (2007).                                                                                       %
\item J. C. Morales, I. Ribas, C. Jordi, {\it Astron. Astrophys.}, {\bf 478}, 507 (2008).                                                             %
\item A. Kraus {\it et al.}, preprint available at http://arxiv.org/abs/1011.2757.                                                                       %
\item I. Baraffe {\it et al.}, {\it Astron. Astrophys.}, {\bf 337}, 403 (1998).                                                                                   %
\item L. Siess, E. Dufour, M. Forestini, {\it Astron. Astrophys.}, {\bf 358}, 593 (2000).                                                                                  %
\item B.-O. Demory {\it et al.}, {\it Astron. Astrophys.}, {\bf 505}, 205 (2009).                                                                          %
\item K. Vida {\it et al.}, {Astron. Astrophys.}, {\bf 504}, 1021 (2009).                                                                                       %
\item P. Demarque {\it et al.}, {\it Astrophys. J. Suppl. Ser.},{\bf 155}, 667 (2004).                                                                                 %
\item  {\"O}. {\c C}akirli, C. {Ibanoglu}, A. {Dervisoglu}, {\it Revista Mexicana de Astronom\'{i}a y Astrof\'{i}sica }, {\bf 46}, 363 (2010) %
\item Funding for this Discovery mission is provided by NASA's Science Mission Directorate.  J. A. C. and D. C. F. acknowledge support for this work was provided by NASA through Hubble Fellowship grants \#HF-51267.01-A and \#HF-51272.01-A awarded by the Space Telescope Science Institute, which is operated by the Association of Universities for Research in Astronomy, Inc., for NASA, under contract NAS 5-26555. J. A. C. is very grateful for helpful discussions with P. Podsiadlowski, S. Rappaport, G. Torres, A. Levine, J. Winn, S. Seager, and T. Dupuy.

\end{enumerate}

\clearpage

Figure 1: Best-fit dynamical-photometric model to the fitted data for KOI-126.  The top eight panels are data observed during the passage of B and C in front of A  from the perspective of the Kepler spacecraft. The bottom two panels are data observed during the passage of B and C behind A (binned by a factor of 10).  In all cases, the solid black line gives the best-fit model.  The inlaid diagrams show, to scale, the orbits of B (green) and C (blue) relative to A (yellow star) at times corresponding to those in the associated time-series.  The dashed orbit is that of the center-of-mass of B and C.  The numbered circles give the locations of B and C at the times indicated by the vertical dotted lines in the associated time-series, corresponding to the numeric index (0--2).  The circles are to scale for the radii of B and C.   The specific values $T_0$ for each respective panel, reading from left to right and top to bottom, are (in BJD) 2455102.815, 2455136.716, 2455170.465, 2455204.267, 2455238.207, 2455271.751, 2455305.713, 2455339.496, 2455259.000, 2455326.506. 

\bigskip

Figure 2: Masses and radii of known low-mass stars with dynamically-estimated masses and radii measured to better than 3\% fractional accuracy ({\it 22, 28}). The black curves correspond to the theoretical stellar isochrones by Baraffe et al.  The dashed, dotted, solid and dash-dotted curves correspond to 1 Gyr, 2 Gyr, 4 Gyr and 5 Gyr solar metallicity isochrones, respectively.  The blue points correspond to CM~Dra  A, B and the red points correspond to KOI-126~B, C. The inset panel corresponds to the region in the larger plot enclosed by the dashed rectangle. 

\clearpage

\begin{figure}
{\large Figure 1}
\centering
\includegraphics[width=6.5in]{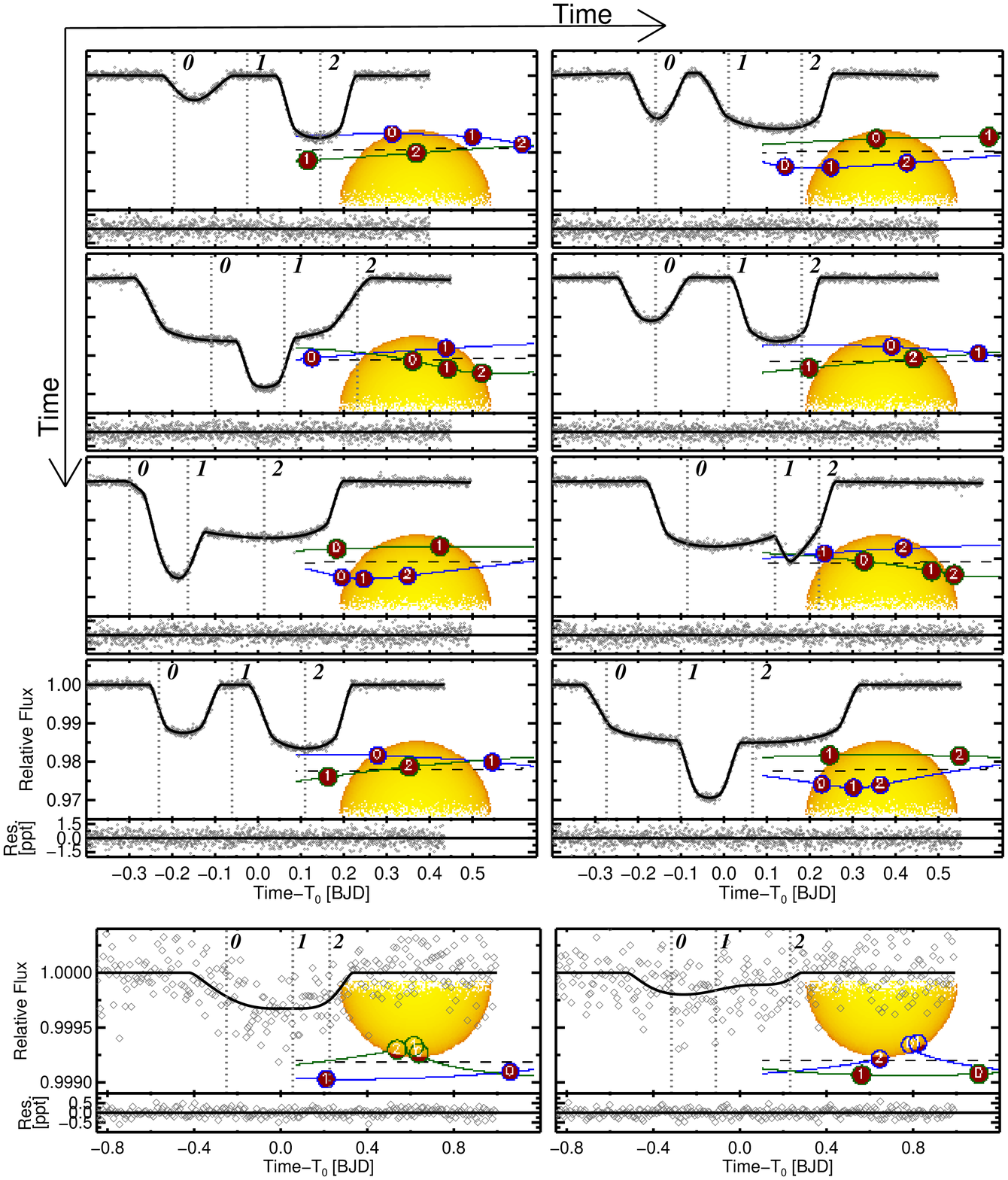}
\end{figure}

\begin{figure}
{\large Figure 2}
\centering
\includegraphics[width=6.5in]{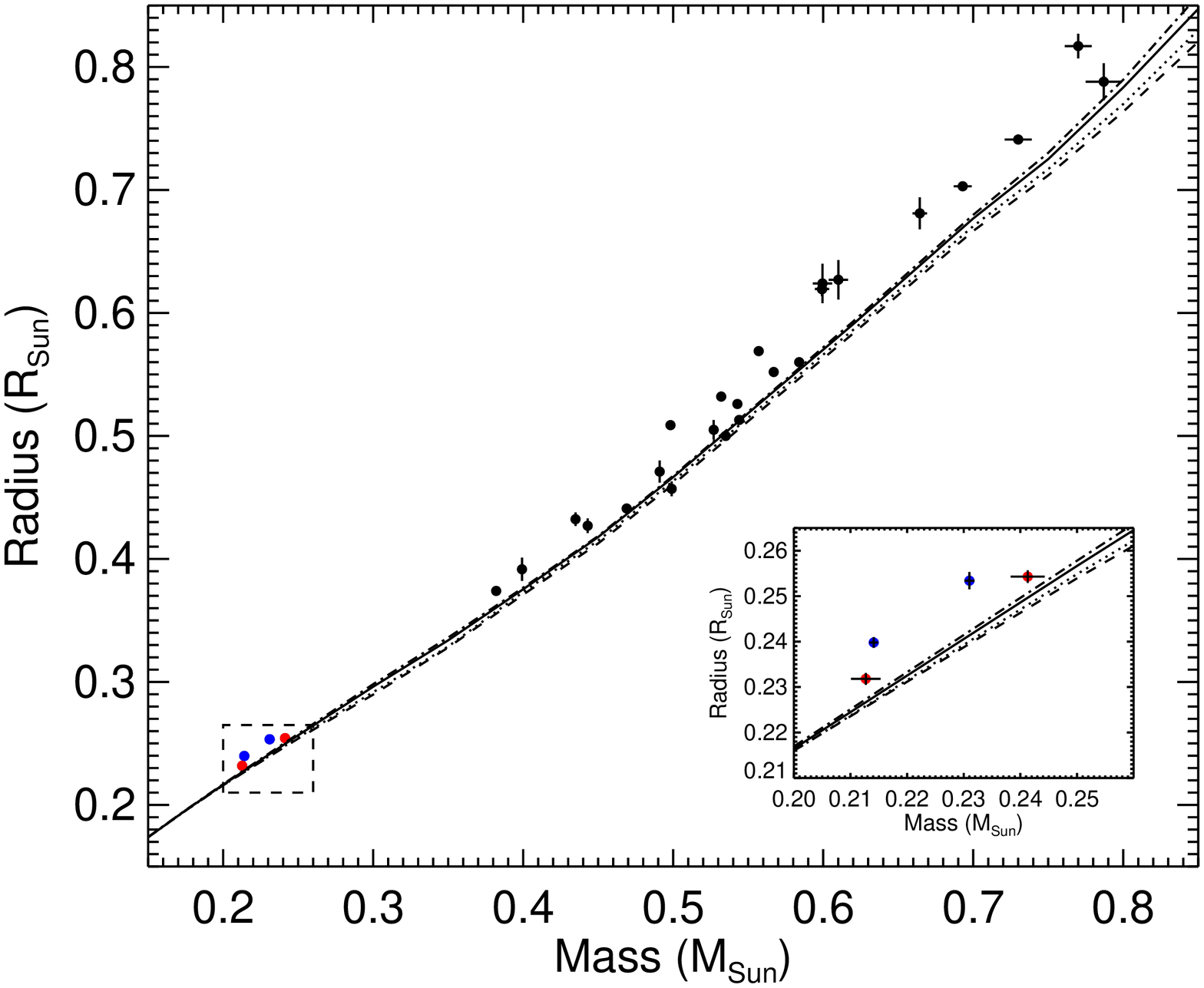}
\end{figure}

\renewcommand{\sp}{\hspace{0.2in}}
\renewcommand{\arraystretch}{1.1}
\begin{table}
\centering
{\footnotesize
\begin{tabular}{|l|l|}
\hline Parameter & Value \\ \hline
Masses & \\ 
	\sp $M_A$ & $1.347\pm0.032$ $M_\odot$ \\   
	\sp $M_B$ & $ 0.2413 \pm 0.0030$ $M_\odot$ \\  
	\sp $M_C$ & $ 0.2127 \pm 0.0026$ $M_\odot$ \\\hline
Radii & \\
	\sp $R_A$ & $2.0254 \pm 0.0098$ $R_\odot$ \\ 
	\sp $R_B$ & $0.2543 \pm 0.0014$ $R_\odot$ \\ 
	\sp$R_C$ & $0.2318 \pm 0.0013$ $R_\odot$ \\\hline 
Average Densities (g cm$^{-3}$) & \\
	\sp$\rho_A$ & $0.2288 \pm 0.0029$\\ 
	\sp$\rho_B$ & $20.70 \pm 0.19$ \\ 
	\sp$\rho_C$ & $24.09 \pm 0.20$ \\\hline
Surface gravities (logarithms in cgs units) & \\
	\sp$\log g_A$ & $3.9547 \pm 0.0069$ \\ 
	\sp$\log g_B$ & $5.0101 \pm 0.0029$ \\ 
	\sp$\log g_C$ & $5.0358\pm 0.0027$ \\\hline
Observed relative fluxes & \\
	\sp$F_A$ & $\equiv 1$\\ 
	\sp$F_B$ & $(3.26 \pm 0.24)\times10^{-4}$\\ 
	\sp$F_C$ & $(2.24 \pm 0.48)\times10^{-4}$ \\\hline
``Outer'' binary [(A, (B, C))] elements on 2,455,170.5 (BJD) & \\
	\sp Period, $P_2$ (day) & $33.9214 \pm 0.0013$ \\ 
	\sp Semi-major Axis, $a_2$ (AU) & $0.2495 \pm 0.0017$ \\
	\sp Eccentricity, $e_2$ & $0.3043 \pm 0.0024$\\ 
	\sp Argument of Periapse, $\omega_2$ &$52.88^\circ\pm0.33$ \\ 
	\sp Mean Anomaly, $M_2$ & $19.87^\circ\pm0.29$ \\ 
	\sp Sky-plane Inclination, $i_2$ & $92.100^\circ\pm 0.016$ \\ 
	\sp Longitude of Ascending Node, $\Omega_2$ & $\equiv 0^\circ$ \\\hline
``Inner'' binary [(B, C)] elements on 2,455,170.5 (BJD) & \\
	\sp Period, $P_1$ (day) & $1.76713 \pm 0.00019$\\
	\sp Semi-major Axis, $a_1$ (AU) & $0.021986 \pm 0.000090 $ \\
	\sp Eccentricity, $e_1$ & $0.02234 \pm 0.00036$ \\ 
	\sp Argument of Periapse, $\omega_1$ & $89.52^\circ \pm 0.42$ \\ 
	\sp Mean Anomaly, $M_1$ &  $355.66^\circ \pm 0.42$\\ 
	\sp Sky-plane Inclination, $i_1$ & $96.907^\circ \pm 0.044$ \\ 
	\sp Longitude of Ascending Node, $\Omega_1$ & $8.012^\circ\pm0.039$\\\hline
Star A parameters from spectroscopy & \\
	\sp Effective Temperature, $T_{\rm eff, A}$ (Kelvin) & $5875\pm100$ \\
	\sp Metallicity, [Fe/H] & $0.15 \pm 0.08$ \\
	\sp Projected Rotational Velocity, $v_A \sin i_A$ (km s$^{-1}$) & $4.6\pm0.9$ \\\hline
\end{tabular}
	}
\caption{Parameters for KOI-126~(A, (B, C)) based on a fit to the Kepler short cadence photometry and sixteen radial velocity measurements. The final section of the table reports the parameters for KOI-126~A based on an analysis of observed spectra having fixed the stellar gravity to that found from the fit to the photometry and RVs. }
\end{table}

\clearpage

\baselineskip16pt

\setcounter{figure}{0}
\setcounter{table}{0}
\renewcommand{\thefigure}{S\arabic{figure}}
\renewcommand{\thetable}{S\arabic{table}}

\paragraph*{\Huge Supporting Online Material}

\paragraph*{Online data}

All data used in this analysis have been made available at {\tt \begin{verbatim}  http://archive.stsci.edu/prepds/kepler_hlsp/\end{verbatim}}

\paragraph*{Spectroscopy}

Six of the sixteen spectra were collected using the Tull Coude
Spectrograph on the 2.7m Harlan J. Smith telescope at the McDonald
Observatory in west Texas, which has a resolving power of R=60,000 and
a wavelength range of ~3750-10000 angstroms. The remaining ten spectra
were obtained using the Tillinghast Reflector Echelle Spectrograph
(TRES; {\it S1}) on the 1.5m Tillinghast Reflector at the Fred
L. Whipple Observatory on Mt. Hopkins, AZ. They were taken with the
medium fiber, corresponding to R=44,000 and a wavelength range of
~3850-9100 angstroms. 

The spectra were extracted and analyzed according to an established
procedure ({\it S2}). We used multi-order
cross-correlations to obtain precise relative velocities separately
for the TRES and McDonald datasets. Spectral orders containing
contaminating atmospheric lines were rejected, along with orders in
the blue with low signal-to-noise and orders in the red with known
reduction problems. In total, we used 18 spectral orders in the
McDonald cross-correlations, covering the wavelength range of
4500-6680 angstroms, and 18 spectral orders in the TRES
cross-correlations, covering the wavelength range of 4580-6520
angstroms. The TRES velocities were shifted to an absolute scale using
the weighted mean offset from the single order velocities and the
known TRES zeropoint (determined from the long term monintoring of IAU
RV standards). The McDonald offset was fit for in the final joint
solution.

We shifted and coadded each dataset and classified the combined
spectra by cross-correlating against the grid of CfA synthetic
spectra, which are based on Kurucz models calculated by John Laird and
rely on a linelist compiled by Jon Morse. Interpolation between
grid-points using the correlation peak heights yields the best-fit
parameters, and the RMS scatter of the results from individual
spectral orders provides an internal error estimate. Because of
degeneracies in the stellar spectrum between $T_{\rm eff}$, $\log g_A$, and [Fe/H],
these parameters are highly correlated, and an error in one parameter
would likely result in systematic errors beyond the internal error
estimates for the other parameters. For this reason, we double the
errors to account for possible systematic effects. With $\log g_A$ fixed
at 3.95, we find $T_{\rm eff}$=5875 $\pm$ 100 K, [Fe/H]=+0.15 $\pm$ 0.08, and
$v_A \sin i$=4.6 $\pm$ 0.9 km/s.

We infer the rotational period of KOI-126 A to be 22$\pm$6 days, which
is super-synchronous.  We note that the observed 17 day periodicity in
the Kepler light curve is comparable to the inferred rotational
period, which suggests a possible association with the rotational
modulation of surface features on KOI-126 A.  The variation is very roughly sinusoidal and out of phase with the eclipse events.  While the period of this variation is almost exactly half the period of the orbit of KOI-126~(B, C) about KOI-126~A, it is unlikely, given the amplitude and phase, that it is associated with the ellipsoidal distortion of A by the tidal field of (B, C) ({\it S3}).

\paragraph*{Dynamical-Photometric Model}

{\it Positions and velocities.} A hierarchical (or Jacobi) coordinate system is used when calculating the positions of the three bodies.  In this system, ${\bf r_1}$ is the position of C relative to B and ${ \bf r_2}$ is the position of A relative to the center of mass of (B,C).  We may specify ${\bf r_1}$ and ${\bf r_2}$ in terms of osculating Keplerian orbital elements (period, eccentricity, argument of pericenter, inclination, longitude of the ascending node, and the mean anomaly: $P_{1,2}$, $e_{1,2}$, $i_{1,2}$, $\omega_{1,2}$, $\Omega_{1,2}$, $M_{1,2}$, respectively).  

Newton's equations of motion, which depend on ${\bf r_1}$, ${\bf r_2}$ and the stellar masses, may be specified for the accelerations ${\ddot{\bf r}_1}$ and ${\ddot{\bf r}_2}$ ({\it S4, S5}).  An additional term may be added to the acceleration of ${\bf r}_1$ due to the post-Newtonian potential of the inner binary ({\it S6}).  These are the only accelerations used in the fit giving the parameters listed in Table 1.  We worked in units such that $G \equiv 1$.

Further perturbing accelerations may be added to the acceleration of ${\bf r}_1$  corresponding to the non-dissipative equilibrium tidal potential between B and C and the potential associated with the rotationally-induced oblate distortion of B and C ({\it S4}).  In this approximation, the axial spins of B and C follow the evolving orbit, staying normal to the orbit and spinning at a rate synchronous with the orbit. Both the accelerations due to tides and rotations depend on ${\bf r}_1$, $M_B$, $M_C$, $R_B$, $R_C$, $k_{2, B}$ and $k_{2,C}$.  The acceleration due to rotation also depends on the angular axial spin rate of both B and C.  The spin rates and apsidal constants are assumed to be the same for both B and C.  We do not model the tidal or rotational distortion of A as their contributions to the total accelerations are negligible ({\it S4}).

A final acceleration due to tidal damping may be added to the acceleration of ${\bf r}_1$ ({\it S5}).  The scale of this acceleration is set by the tidal dissipation efficiency, $Q$.  This acceleration is negligible for reasonable values of $Q$ ($> 100$). 

We used an implementation of the Bulirsch-Stoer  algorithm ({\it S7}) to numerically integrate the coupled first-order differential equations for $\dot{\bf r}_{1,2}$ and ${\bf r}_{1,2}$ in order to determine ${\bf r}_{1,2}$ and their temporal derivatives at any given time.  The maximum step size in the integrator was $P_1/1000 \approx 3$ min. 

The Jacobi coordinates (${\bf r_1}$ and ${\bf r_2}$ and their derivatives) may be transformed into spatial coordinates of the three bodies, relative to barycenter.

{\it Radial velocity of KOI-126 A.} The RV data for KOI-126 A were compared directly to the results of the numerical algorithm after applying a systematic offset associated with peculiar and bulk Galactic motion and an additional offset between the McDonald and TRES spectra to account for calibration error.  The systematic offset was measured to be $\gamma = -27.278 \pm 0.024$ km~s$^{-1}$ and the additional offset was measured to be $\gamma' = -0.26 \pm 0.17$ km~s$^{-1}$. 

{\it Correcting for the finite speed of light.} The positions of the three stars are projected to the location of the barycentric plane (i.e., the plane parallel to the sky-plane that includes the barycenter of KOI-126 and that is normal to the line-of-sight) at a time $t_0$ in order to correct for the delay resulting from the finite speed of light and the motion of the barycenter of KOI-126 along the line-of-sight.  In detail, for each star and for a given observation (or ``clock'') time $T$, a secant line root finding algorithm is used to solve $d(t_r)-c ( T-t_r)-\gamma (T-t_0)= 0$ for the retarded time, $t_r$, where $d(t)$ is the star's distance at a time $t$ from the barycentric plane, $\gamma$ is the radial velocity of the barycenter and $c$ is the speed of light.  The observed coordinates of each star at the clock time $T$ are those found at each individual body's retarded time.  For $\gamma \equiv 0$, the variation of the difference between the clock time and the retarded time is as much as a few minutes, in amplitude.  We note that the light-time effects set the physical scale of the system and allow for the determination of absolute masses and radii ({\it S8}).  Our code permits the use of a non-zero barycentric radial velocity $\gamma$ which may be measured from the radial velocity data. The effect of this parameter is to shift the model parameters with a fractional difference similar to $\gamma/c \approx 10^{-4}$.  The only parameters for which this effect is relevant are the orbital periods of the inner and outer binary which are adjusted by $\approx 2.5$ and $\approx 0.88$ sigma, respectively, relative to the periods measured with $\gamma \equiv 0$.  All remaining parameters are adjusted by less than one tenth of their respective one sigma uncertainty.  Given this fact and the possibility of a systematic bias in the velocity zeropoint, we opted to report (in Table 1) the parameters found assuming $\gamma \equiv 0$.

{\it Photometric model.} The sky-plane projected 2D positions of all three objects were used as inputs to a light curve generating algorithm.  

All stars were assumed to be spherical.  Additionally, the radial brightness profile of KOI-126 A was modeled as $I(r)/I(0) = 1-u_1 (1-\sqrt{1-r^2})-u_2 (1-\sqrt{1-r^2})^2$ where $r$ is the projected distance from the center of A, normalized to the radius of A, and $u_1$ and $u_2$ are the two quadratic limb-darkening parameters ({\it S9}).  The fluxes of B and C were specified relative to the flux of A.   The sum of the fluxes was normalized to unity.  The limb-darkening coefficients for A were found to be $u_1 = 0.39 \pm 0.03$ and $u_2 = 0.22 \pm 0.04$.  These values agree with theoretical expectations for stars with temperatures, gravities and metallicities similar to those found for KOI-126 A ($0.3552 < u_1 < 0.4009$ and $0.2553 < u_2 < 0.2869$ for $5750 K< T_{\rm eff} < 6000 K$, $\log g_A \equiv 4.0$ and $0.10 < $[M/H]$ < 0.20$,  {\it S10}).  We investigated various limb-darkened profiles for B and C according to the same model, however, any physically-valid choice of parameters gave similar fit values within the quoted uncertainties.

For eclipses in which there are no three-way alignments, the loss of light is computed using an analytic prescription that depends on object separations, radii and relative fluxes ({\it S11}). The total loss of light at a given time is the sum of the losses associated with the eclipse of A by B or C, the eclipse of B or C by A, or the eclipses between B and C.   When a three-way alignment occurs, the total loss of light is not given as a superposition of analytically-defined overlaps.  In this case, the loss of light is computed numerically by integrating over the sky-plane.  

\paragraph*{Determining best-fit parameters, covariances and uncertainties}

{\it Fitting parameters.} There were 23 free parameters in the final fit to all available data: the three masses ($M_A$, $M_B$, $M_C$), five orbital elements of the outer binary at $t_0 = 2,455,170.5$ (BJD) ($P_2$, $e_2$, $\omega_2$, $i_2$, $M_2$), six orbital elements of the inner binary at $t_0$ ($P_1$, $e_1$, $\omega_1$, $i_1$,$\Omega_1$,$M_1$),  the radius of A ($R_A$), the relative radii of B and C ($R_B/R_A$, $R_C/R_A$), the relative fluxes of B and C ($F_B/F_A$, $F_C/F_A$), the two limb-darkening parameters of A ($u_1$, $u_2$), the systematic offset to the radial velocity of A ($\gamma$) and a systematic offset between TRES and McDonald RVs ($\gamma'$).  The longitude of the ascending node of the outer binary is unconstrained and, for simplicity, has been fixed to $\Omega_2 = 0^\circ$.  The longitude of the inner binary is measured relative to this orientation and does not reflect the true value.  

All short cadence data were initially fit with a multiplicative correction that was quadratic in time, to account for out-of-eclipse long-wavelength variability.  The parameters describing these corrections correlated very weakly with the remaining 23 parameters and were therefore fixed to their best-fit values to reduce computation time.

{\it Best-fit parameters, errors and covariances.} The data were fitted using an implementation of the Levenberg-Marquardt (L--M) algorithm ({\it S12, S13}).  The L--M algorithm minimizes the sum of the deviates squared ($\chi^2$) by adjusting the model parameters.  In addition to determining the best-fit parameters, the L--M algorithm also returns the covariance matrix of the fitted parameters, approximating the $\chi^2$--surface as quadratic in those parameters. For properly estimated measurement uncertainty, the square root of the diagonal elements gives the formal 1--sigma statistical errors in the parameters.  The statistical error in any derived parameter may be estimated using the full covariance matrix.  %

The measurement uncertainties for the photometric data were set equal to the root-mean-square deviation of the best-fit residuals for each short cadence event as plotted in Fig. 1.  The best-fit residuals were observed to show very little temporal correlation.  The best-fit solution had a reduced-$\chi^2$ of $1.00012$ for $17370$ degrees of freedom.  The $\chi^2$ contribution from the RV data alone was $13.25$.

  We note that while the masses $M_B$ and $M_C$ are known to no better than 1.5\%, their ratio is known much more precisely: $M_B/M_C=1.1349\pm0.0012$.  The ratio of radii are also well-determined: $R_B/R_A =  0.12558\pm0.00021$ and $R_C/R_A = 0.11445\pm0.00015$.  Also, the sum of the radii of B and C are known relatively better than their difference -- $R_B+R_C = 0.4861\pm0.0026$  $R_\odot$ and $R_B-R_C = 0.02254\pm0.00026$ $R_\odot$.  The sum and differences of the masses of B and C have an analogous structure -- $M_B+M_C =  0.4540\pm0.0057$ $M_\odot$ and $M_B-M_C = 0.02868\pm0.00042$ $M_\odot$.
  
 We executed the L--M algorithm, starting from random positions in parameter space, many hundreds of times before finding the final solution.  
 From this experience we note that a large number of local maxima ($\chi^2$ minima) populate the likelihood landscape. In particular, solutions with inner binary periods belonging to a discrete set of aliases of the reported inner binary period satisfying $P_1' = P_2/(P_2/P_1 + n)$, for integers $n$, show qualitatively similar transit events.  These solutions yield residual correlated structure, significantly higher $\chi^2$ and physically implausible stellar parameters for KOI-126 A.  Moreover, these solutions do not predict the observed occultations or inner binary eclipses.  A truncated Markov chain Monte Carlo search of the likelihood surface surrounding the adopted solution uncovered no larger extrema nor any local extrema with interestingly-low $\chi^2$.  
 
 \paragraph*{Tidal Dissipation and Timescales}
 
Tidal dissipation is expected to modify the orbital elements of the close binary and the spins of its components ({\it S14}).  In isolation from KOI-126~A, the orbital eccentricity of KOI-126 (B,C) should decrease with an exponential timescale $\sim 1$ Gyr if $Q/k_2 = 10^7$ for each of B and C, where $Q$ is the standard tidal quality factor ({\it S15}).  In the presence of KOI-126 A, the eccentricity would damp not to zero, but to a small fixed value, with the eccentricities of the close binary and third star aligned and precessing at the same rate ({\it S16}).  However, the observed eccentricity ($\langle e_1 \rangle \approx 0.01$) has not yet damped to that state (Figure S5).  Moreover, if the system endured more than $\sim3$ exponential damping timescales, then $e_1>0.25$ in the past, in violation of dynamical stability ({\it S17}).  Together with estimates of $k_2\simeq0.1$, these arguments bound $Q \gtrsim 10^6$.  

The nearly coplanar configuration of the triple system (9.2$^\circ$, oscillating by 0.4$^\circ$ on each orbit of the third body) rules out significant eccentricity limit cycles ({\it S18}) or Kozai oscillations ({\it S19}) in the close binary.  Aside from eccentricity damping, in isolation from KOI-126 A, the spins of KOI-126 B and C would damp quickly ({\it S14}) ($\sim 10^8$ years) to an orbit-aligned and orbit-synchronized state.  The presence of KOI-126 A causes the KOI-126 (B,C) orbit to precess faster than the natural spin precession of the component stars, so the spin vectors would damp to nearly the precession axis instead ({\it S20}).  Although our observations are not currently sensitive to such spin-orbit misalignment, its long-term effect is to damp the mutual inclination ({\it S19}) on a timescale several to ten times longer than the eccentricity-damping timescale.

\clearpage

\begin{figure}
\centering
\includegraphics[width=6.5in]{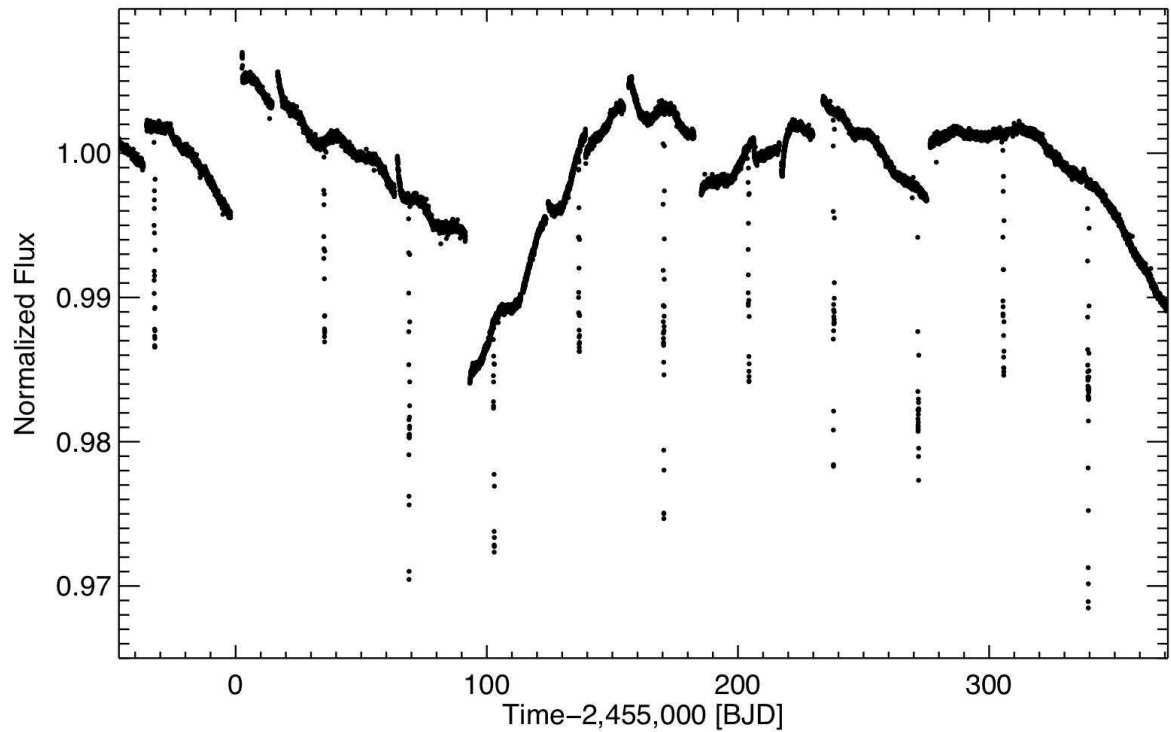}
\caption{Long cadence Kepler light curve for KOI-126 A showing data from Quarters 1--5.  The plotted data have been normalized by a constant value but are otherwise the unprocessed product of Kepler aperture photometry.  The discontinuities correspond to quarter breaks at which point the Kepler spacecraft is rolled and KOI-126 is then observed at different pixels on the photometer. }
\end{figure}

\begin{figure}
\centering
\includegraphics[width=6.5in]{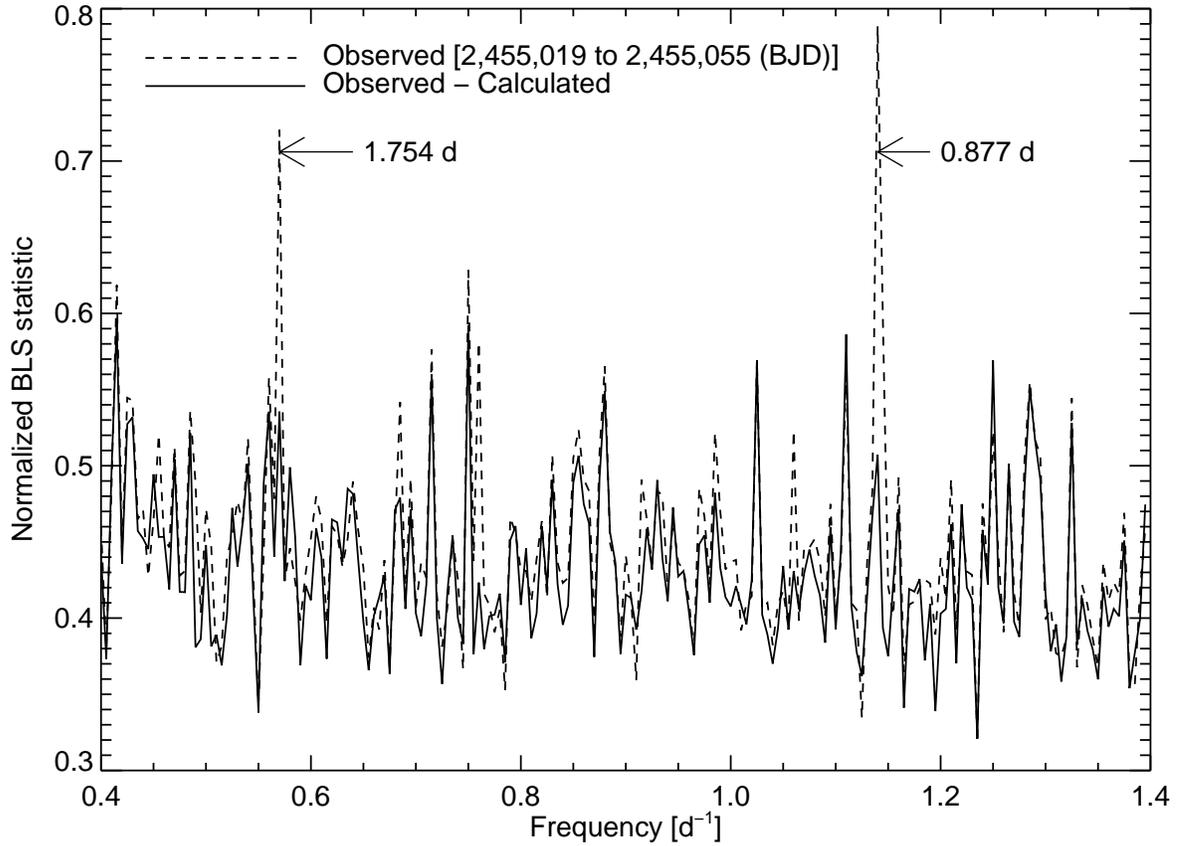}
\caption{Results of a search for eclipses between KOI-126 B and C.  Plotted is the ``spectrum'' (dashed curve) from the Box-Least-Squares (BLS) search algorithm ({\it S21}) indicating periodic dips in the Kepler photometry over 36 continuous days of observation, after excluding the prominent transit eclipses and de-trending by dividing by a 10 day moving average.  The peak period corresponds to approximately half the orbital period of KOI-126 B and C.  The solid curve shows the result of the BLS algorithm after subtracting the best-fit photometric solution.  A similar BLS search of data spanning an identical interval, observed 100 days later than this data, does not show any structure, indicating the eclipses had stopped. }
\end{figure}

\begin{figure}
\centering
\includegraphics[width=6.5in]{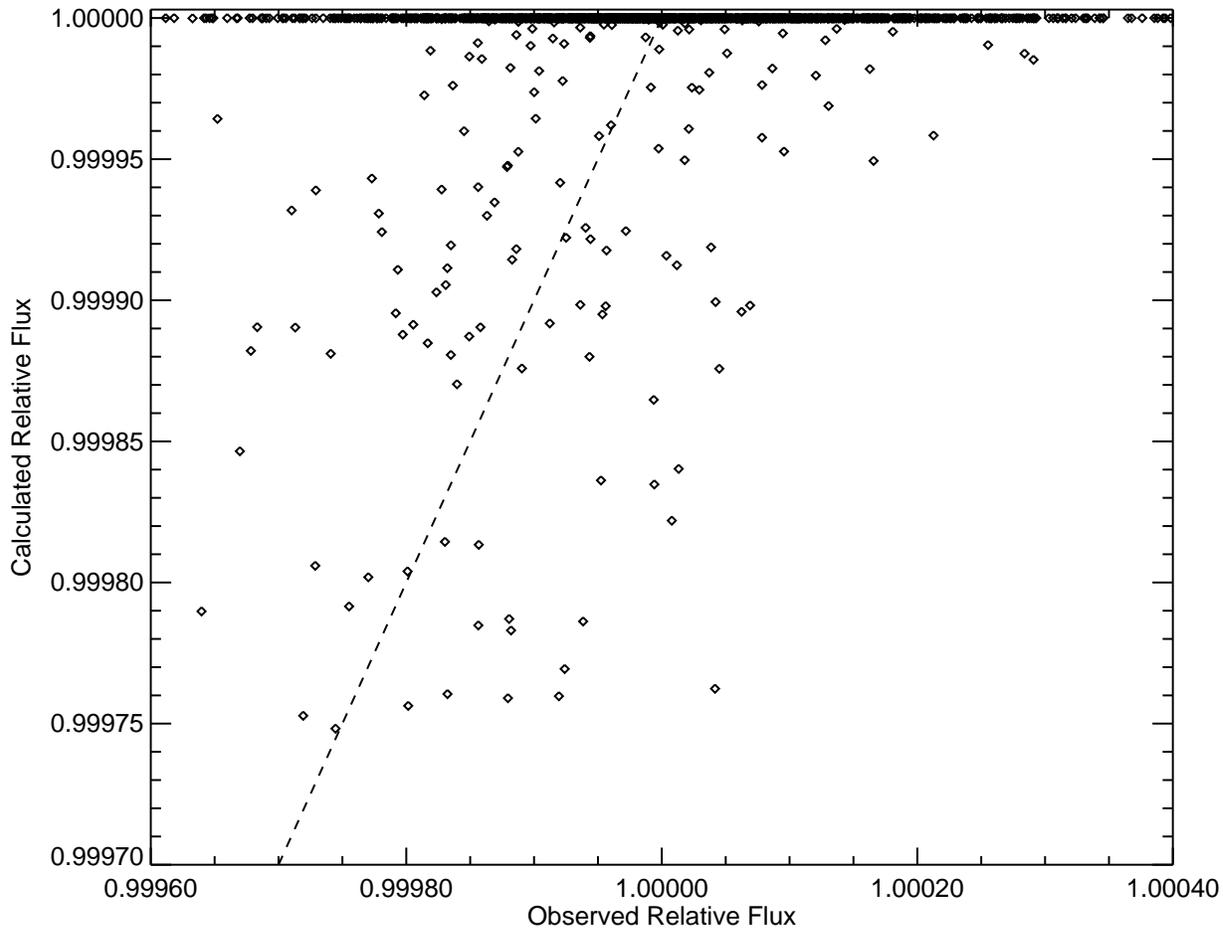}
\caption{Comparison of the observed and calculated light curve for the eclipses between KOI-126 B and C.  The observed data are those utilized in the BLS visually represented in Fig. S2.  The calculated model is that using the best-fit parameters based upon a fit to the data in Fig. 1.  The dashed line is 1:1.}
\end{figure}

\begin{figure}
\centering
\includegraphics[width=6.5in]{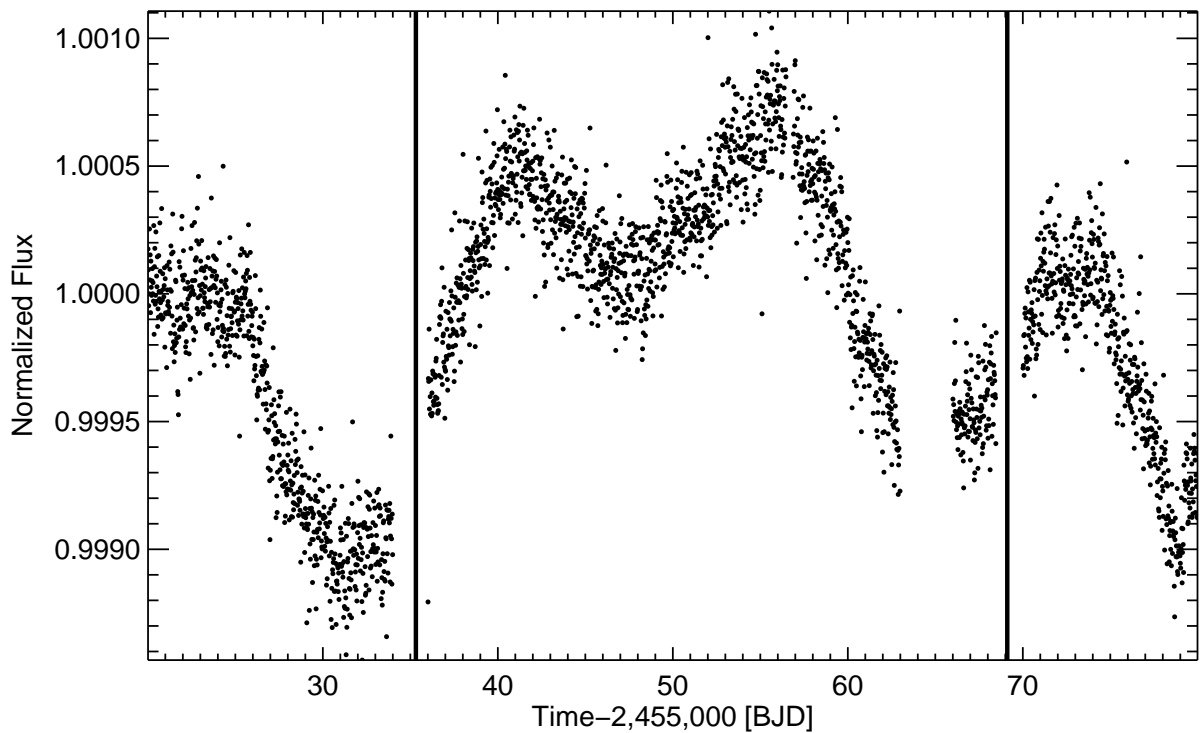}
\caption{Out-of-eclipse variability of KOI-126.  Plotted is a portion of the full Kepler light curve for KOI-126 showing a $\approx 17$ day modulation.  The solid vertical lines indicate the location in time of two transit events. }
\end{figure}

\begin{figure}
\centering
\includegraphics[width=6.5in]{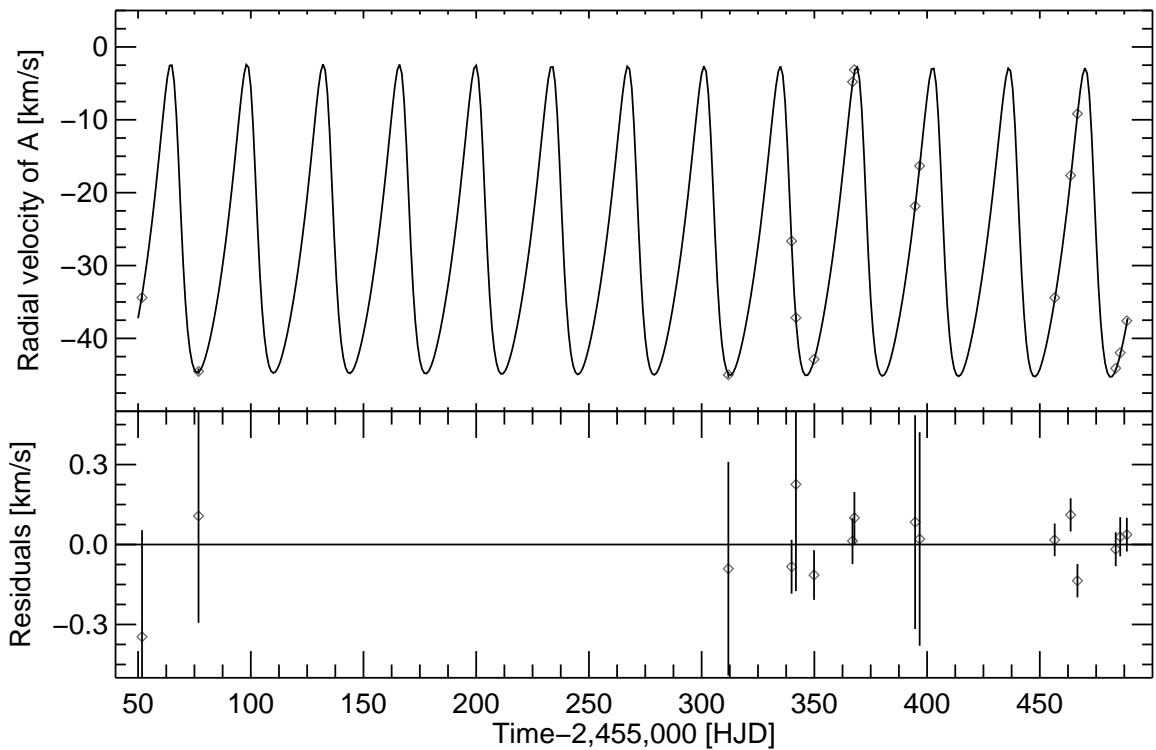}
\caption{Radial velocity data for KOI-126 A.  The solid black curve gives the best-fit model to the data.  The data with the larger ($\approx 0.5$ km s$^{-1}$) errors corresponds to McDonald measurements while the more precise data correspond to TRES measurements.}
\end{figure}

\begin{figure}
\centering
\includegraphics[width=6.5in]{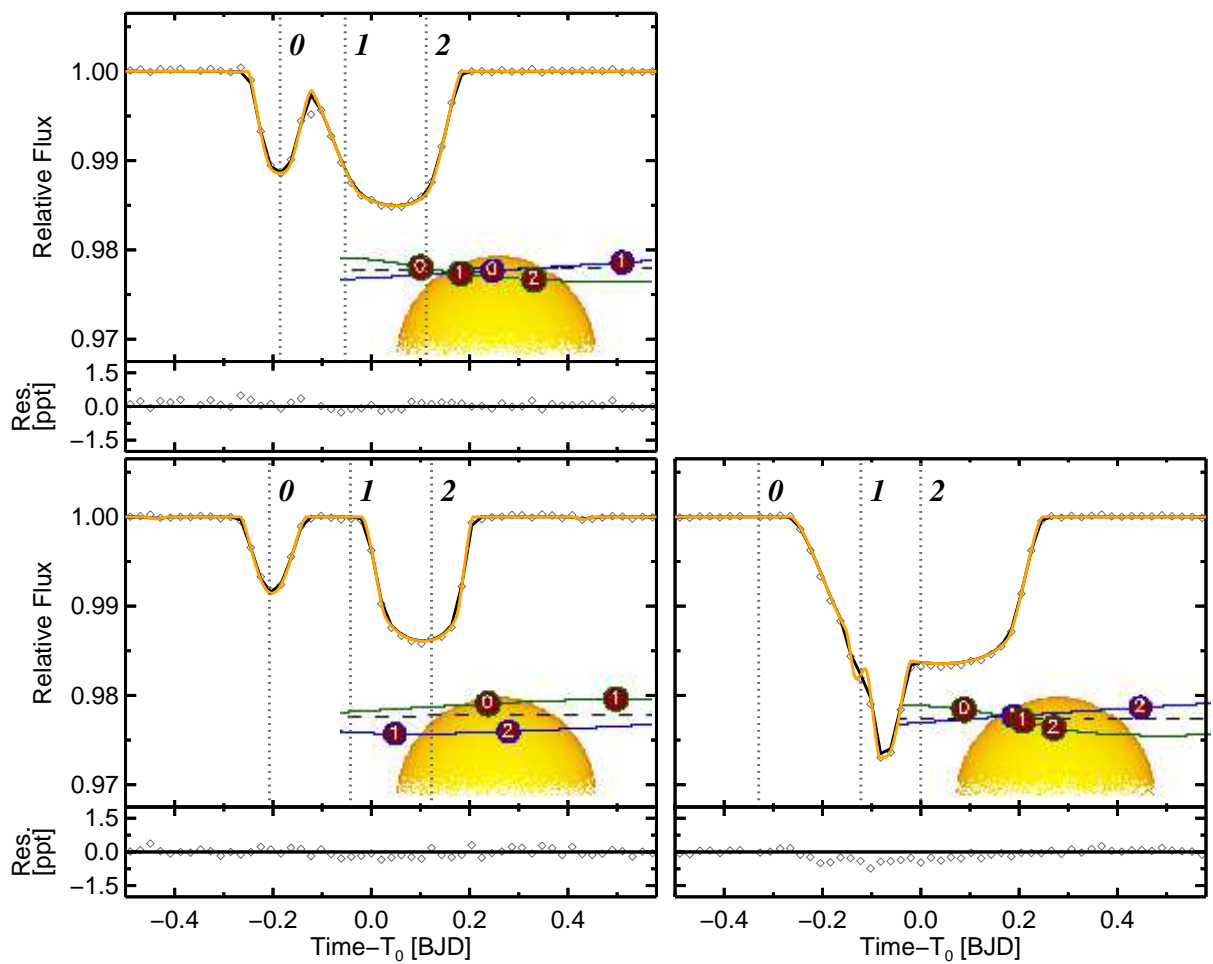}
\caption{Long cadence data for KOI-126 corresponding to the passage of KOI-126 (B,C) in front of KOI-126 A.  The orange curve is the continuous model corresponding to the integrated best-fit model (in black).  Refer to the caption of Fig. 1 for additional figure details.  The specific values $T_0$ for each respective panel, reading from left to right and top to bottom, are (in BJD) 2454967.760, 2455035.316, 2455069.113. }
\end{figure}

\begin{figure}
\centering
\includegraphics[width=6.5in]{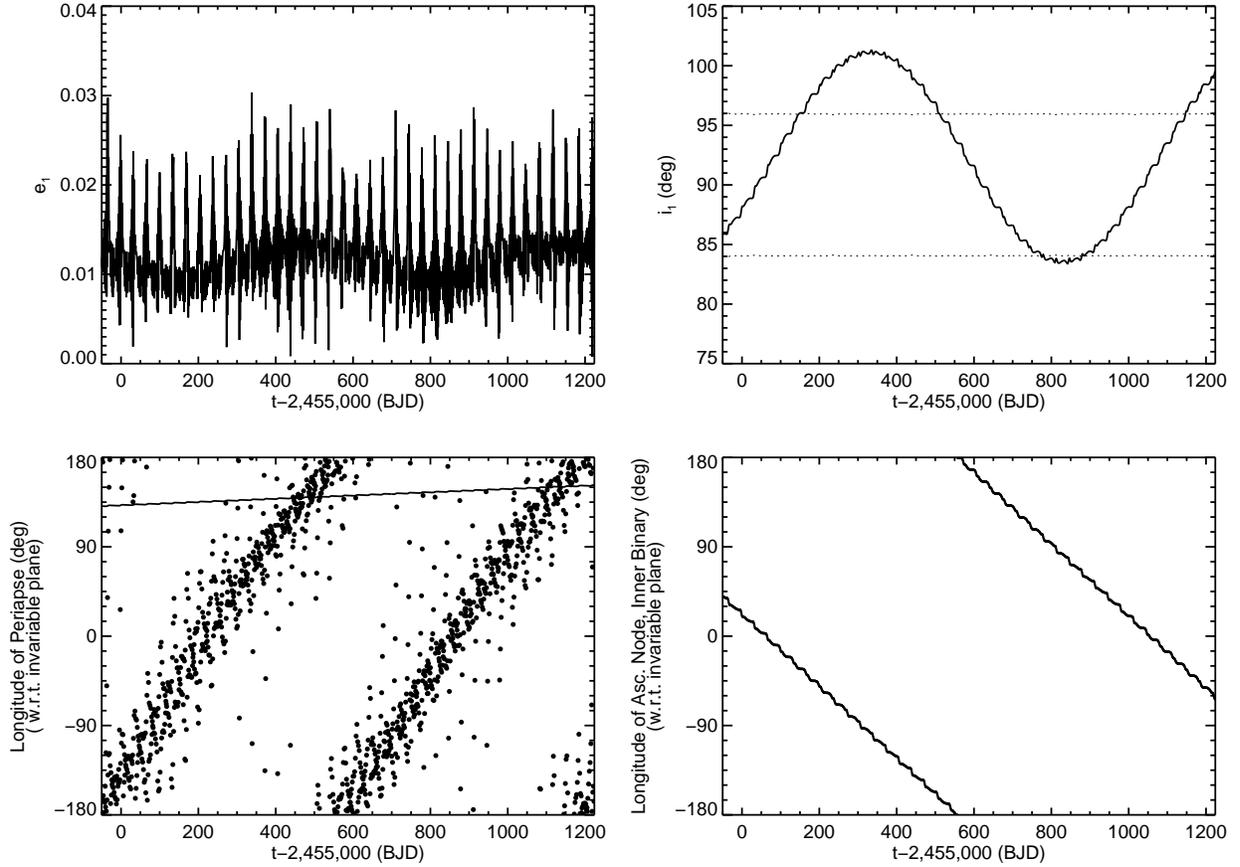}
\caption{Evolution of the Keplerian orbital elements of KOI-126.  The upper left panel shows the evolution of the eccentricity of the  inner binary (composed of KOI-126~B and C).  The period of the secular variation in eccentricity is $\sim 650$ days. The large jumps in eccentricity correspond to periastron passage in the outer binary.  The upper right panel shows the evolution of the inner binary inclination.  The dashed lines bound the region in inclination for which eclipses of the inner binary will occur.  The bottom left panel shows the evolution of the argument of periapse for the inner binary (dots) and for the outer binary [composed of the center of mass of KOI-126~(B, C) and KOI-126~A; solid line] relative to the invariable plane that is perpendicular to the total angular momentum.  The apsidal alignment of the inner and outer binaries corresponds to a maximum in the secular component of the inner binary eccentricity.  The bottom right panel shows the evolution of the nodal longitude of the inner binary, relative to the invariable plane.  The period secular variation of the inclination and the node is $\sim 950$ days.  The span of time covered in the plots corresponds to the nominal 3.5 year Kepler mission.  Times from 2,454,950 (BJD) to 2,455,370 (BJD) cover the Kepler observations included in this work.}
\end{figure}

\clearpage

\begin{table}
\centering
{
\begin{tabular}{lll}
\hline Time (BJD) & Radial velocity of A (km s$^{-1}$) & RV Error (km s$^{-1}$) \\ \hline
      2455339.898438 &  -26.651 &    0.101 \\ 
      2455349.863281 &  -42.856 &    0.093 \\ 
      2455366.925781 &   -4.787 &    0.087 \\ 
      2455367.765625 &   -3.123 &    0.097 \\ 
      2455456.621094 &  -34.408 &    0.061 \\ 
      2455463.695312 &  -17.632 &    0.062 \\ 
      2455466.714844 &   -9.171 &    0.063 \\ 
      2455483.679688 &  -44.097 &    0.064 \\ 
      2455485.667969 &  -41.976 &    0.073 \\ 
      2455488.613281 &  -37.594 &    0.063 \\ 
      2455051.753906 &  -34.411 &    0.400 \\ 
      2455076.828125 &  -44.535 &    0.401 \\ 
      2455311.859375 &  -44.997 &    0.401 \\ 
      2455341.812500 &  -37.164 &    0.401 \\ 
      2455394.703125 &  -21.844 &    0.401 \\ 
      2455396.714844 &  -16.319 &    0.401 \\
\end{tabular}
	}
\caption{Radial velocity measurements for KOI-126~A.  The adopted error is the sum in quadrature of the internal error and the estimated error attributed to instrument stability. }
\end{table}

\clearpage

{\bf References and Notes}
\renewcommand{\labelenumi}{S\arabic{enumi}.}
\begin{enumerate}
\item G. F\H{u}r\'esz, Ph.D.~thesis, University of Szeged, Hungary (2008).
\item L. A. Buchhave {\it et al.}, {\it Astrophys. J}, {\bf 720}, 1118 (2010).
\item S. L. Morris, {\it Astrophys. J.}, {\bf 295}, 143 (1985).                                                                                                           %
\item S. Soderhjelm, {\it Astron. Astrophys.}, {\bf 141}, 232 (1984).                                                                    			%
\item R.~A. Mardling, D.~N.~C. Lin, {\it Astrophys. J.}, {\bf 573}, 829 (2002).								%
\item M.~H. Soffel, Relativity 																	%
in Astrometry, Celestial Mechanics and Geodesy, XIV, 
Springer-Verlag Berlin Heidelberg New York. (1989)
\item W. H. Press {\it et al.}, {\it Numerical Recipes in C++} (2007).
\item D. Ragozzine, M.J. Holman, {\it Astrophys. J.}, preprint available at http://arxiv.org/abs/1006.3727.                     %
\item A. Claret, {\it Astron. Astrophys.}, {\bf 363}, 1081 (2007)
\item D. Sing, {\it Astron. Astrophys.}, {\bf 510}, A21 (2010).
\item K. Mandel, E. Agol, {\it Astrophys. J.}, {\bf 580}, L171 (2002).
\item C. B. Markwardt, {\it Astron. Data Analysis Software and Systems XVII ASP Conference Series}, {\bf 411}, 251 (2009).
\item J. Mor\'{e}, {\it Numerical Analysis}, {\bf 630}, 105 (1978). 
\item J.-P. Zahn, {\it Astron. Astrophys.}, {\bf 57}, 383 (1977).                                                                                                    %
\item B. Jackson, R. Greenberg, R. Barnes, {\it Astrophys. J.}, {\bf 678}, 1396 (2008).
\item R. A. Mardling, {\it Mon. Not. R. Astron. Soc.}, {\bf 382}, 1768 (2007).                                                                        %
\item P. Eggleton, L. Kiseleva, {\it Astrophys. J}, {\bf 455}, 640 (1995).                                                                                    %
\item R.~A. Mardling, {\it Mon. Not. R. Astron. Soc.}, {\bf 407}, 1048 (2010). 
\item D. Fabrycky, S. Tremaine, {\it Astrophys. J.}, {\bf 669}, 1298 (2007).                                                                              %
\item S. Peale, {\it Astron. J.}, {\bf 74}, 483 (1969).                  											%
\item G. Kov{\'a}cs, S. Zucker, T. Mazeh, {\it Astron. Astrophys.}, {\bf 391}, 369 (2002). 						%

\end{enumerate}										%

\end{document}